\documentclass[10pt,twocolumn,showpacs,amsmath,amssymb,floatfix,superscriptaddress]{revtex4-2}

\usepackage[T1]{fontenc}
\usepackage{graphicx}
\usepackage{dcolumn}
\usepackage{bm}
\usepackage{color}
\usepackage{txfonts}
\usepackage{microtype}
\usepackage{hyperref}
\usepackage[english]{babel}
\usepackage{slashed}
\usepackage{gensymb}
\usepackage{epsfig}
\usepackage[normalem]{ulem}
\usepackage{amsmath}
\usepackage{array}
\usepackage{booktabs}
\usepackage{nccmath}
\allowdisplaybreaks[4]

\begin{document}
\renewcommand{\figurename}{FIG}	
	
\title{On-chip high energy photon radiation source based on microwave-dielectric undulator}

\author{Fuming Jiang}  \affiliation{State Key Laboratory of High Field Laser Physics and CAS Center for Excellence in Ultra-intense Laser Science, Shanghai Institute of Optics and Fine Mechanics (SIOM), Chinese Academy of Sciences (CAS), Shanghai 201800, China}
\affiliation{University of Chinese Academy of Sciences, Beijing 100049, China}	

\author{Xinyu Xie}  	\affiliation{State Key Laboratory of High Field Laser Physics and CAS Center for Excellence in Ultra-intense Laser Science, Shanghai Institute of Optics and Fine Mechanics (SIOM), Chinese Academy of Sciences (CAS), Shanghai 201800, China}
\affiliation{University of Chinese Academy of Sciences, Beijing 100049, China}	

\author{Chengpu Liu}\email{chpliu@siom.ac.cn}
\affiliation{State Key Laboratory of High Field Laser Physics and CAS Center for Excellence in Ultra-intense Laser Science, Shanghai Institute of Optics and Fine Mechanics (SIOM), Chinese Academy of Sciences (CAS), Shanghai 201800, China}
\affiliation{University of Chinese Academy of Sciences, Beijing 100049, China}	

\author{Ye Tian}\email{tianye@siom.ac.cn}
\affiliation{State Key Laboratory of High Field Laser Physics and CAS Center for Excellence in Ultra-intense Laser Science, Shanghai Institute of Optics and Fine Mechanics (SIOM), Chinese Academy of Sciences (CAS), Shanghai 201800, China}
\affiliation{University of Chinese Academy of Sciences, Beijing 100049, China}

\begin{abstract}
A new on-chip light source configuration has been proposed, which utilizes the interaction between microwave and a dielectric nanopillar array to generate a periodic electromagnetic near field, and applies periodic transverse acceleration to relativistic electrons to generate high-energy photon radiation. Here the dielectric nanopillar array interacting with microwave acts as the electron undulator, in which the near field drives electrons to oscillate. When an electron beam operates in this nanopillar array in this light source configuration, it is subjected to a periodic transverse near-field force, and will radiate X-ray or even $\gamma$-ray high energy photons after a relativistic frequency up-conversion. Compared with the laser-dielectric undulator based on the interaction between strong lasers and nanostructures to generate a plasmonic near field, this configuration is less prone to damage during operation.
\end{abstract}

\maketitle
\section{Introduction}\label{introduction}
Traditional solid and gas lasers cannot directly generate X-Ray or $\gamma$-Ray. In addition to decay radiation sources, high-energy electronic radiation can effectively generate such high-energy photons. For example, using electrons emitted by electron guns to bombard targets can generate X-ray through bremsstrahlung, or injecting relativistic electrons into undulators can also generate soft/hard X-ray. The former is the oldest X-ray generation plan in history, widely used in medicine, industrial production, and other fields, while the latter is free electron laser (FEL), currently one of the most cutting-edge topics. However, X-ray generated by bremsstrahlung have some defects, such as large divergence angle and poor coherence. And traditional FEL requires a large array of undulatory magnets, this large volume of strong magnetic system needs extremely short (micrometer level) oscillation period, so it is difficult to establish such a device. If we want to use FEL system to generate X-ray or even Gamma-ray, there are two direct ways: (1) we can accelerate the free electrons to extreme relativistic region, so the shortwave radiation can be generated by these high-energy electrons via  relativistic frequency up-conversion effect; (2) We can also build  a longer undulator system to generate high-order harmonics and then amplify the output. Both of these plans will make FEL systems larger and more expensive, and such complex precision systems will also reduce their reliability and increase the difficulty of construction and maintenance \cite{1}.

To solve the above problems, many scientists have taken different approaches and attempted non-magnetic array undulator configurations. Elias proposed the concept of "electromagnetic wave undulator" in 1979, pointing out that electromagnetic waves can interact with free electrons to generate a wave period much smaller than a magnetic array does \cite{2}. The most intuitive principle is the inverse Compton scattering (ICS) gamma laser, where electrons interact with high-energy laser pulses and undergo transverse acceleration under the laser electromagnetic field, then stimulating radiation. At this point, the oscillation period of the electron is close to the oscillation period of the laser pulse, which is much shorter than the traditional undulator period. From a quantum perspective, it can be understood that a free electron undergoes ICS with a low-energy photon, generating a high-energy photon. MIT proposed an ICS gamma laser system based on a linear accelerator in 2009 \cite{3}. In 2012, K. Ta Phuoc \textit{et al}. proposed a "fully optical Compton gamma light source", where the generation, acceleration, and oscillation of free electrons are all achieved by a single light source \cite{4}. With the development of Laser Wake Field Acceleration (LWFA), it has been found that when electrons are accelerated inside plasma bubbles, generated by ponderomotive force, they will produce cyclotron betatron radiation, with significantly shorter oscillation periods than solid undulators. Electrons, without extremely high energy, can also produce strong X-rays \cite{5,6}.

In addition, the concept of "on-chip FEL light source" is also one of the cutting-edge types of research in recent years. There are two technical routes in this plan: (1) surface plasmon polaritons (SPPs) amplification, (2) and micro/nano-structure modulation of electrons. The former includes on-chip light sources that interact with free electrons and SPPs \cite{7,8}, while the latter includes Smith Purcell radiation generated by the interaction between electrons and micro/nano gratings \cite{9,10,11},  and periodic near field generated by micro periodic structures effect electron beams \cite{12,13,14,15}. Although on-chip light sources based on Smith Purcell radiation can also produce X-rays \cite{16}, their directionality and monochromaticity are generally inferior to near-field undulators.

With the improvement of micro/nano-processing technology, micro/nano-structures with periodic of tens of nanometers can be prepared. By using appropriate external pumping, near field can be generated around these periodic micro/nano-structures. When an electron beam passes through the structure, it is subjected to the near field, generating periodic acceleration and radiation. 

In this letter, we propose a model that utilizes the interaction between microwaves and nano-sized structures to generate quasi-static periodic near field, thereby inducing electron oscillations to generate X-Ray or even $\gamma$-Ray, and use simulation to obtain the radiation’s characters of the device.

\section{PERIODIC NEAR-FIELD CHANNEL}\label{theory}
\begin{figure}[h]
	\setlength{\abovecaptionskip}{0.6cm}
	\centering\includegraphics[width=1\linewidth]{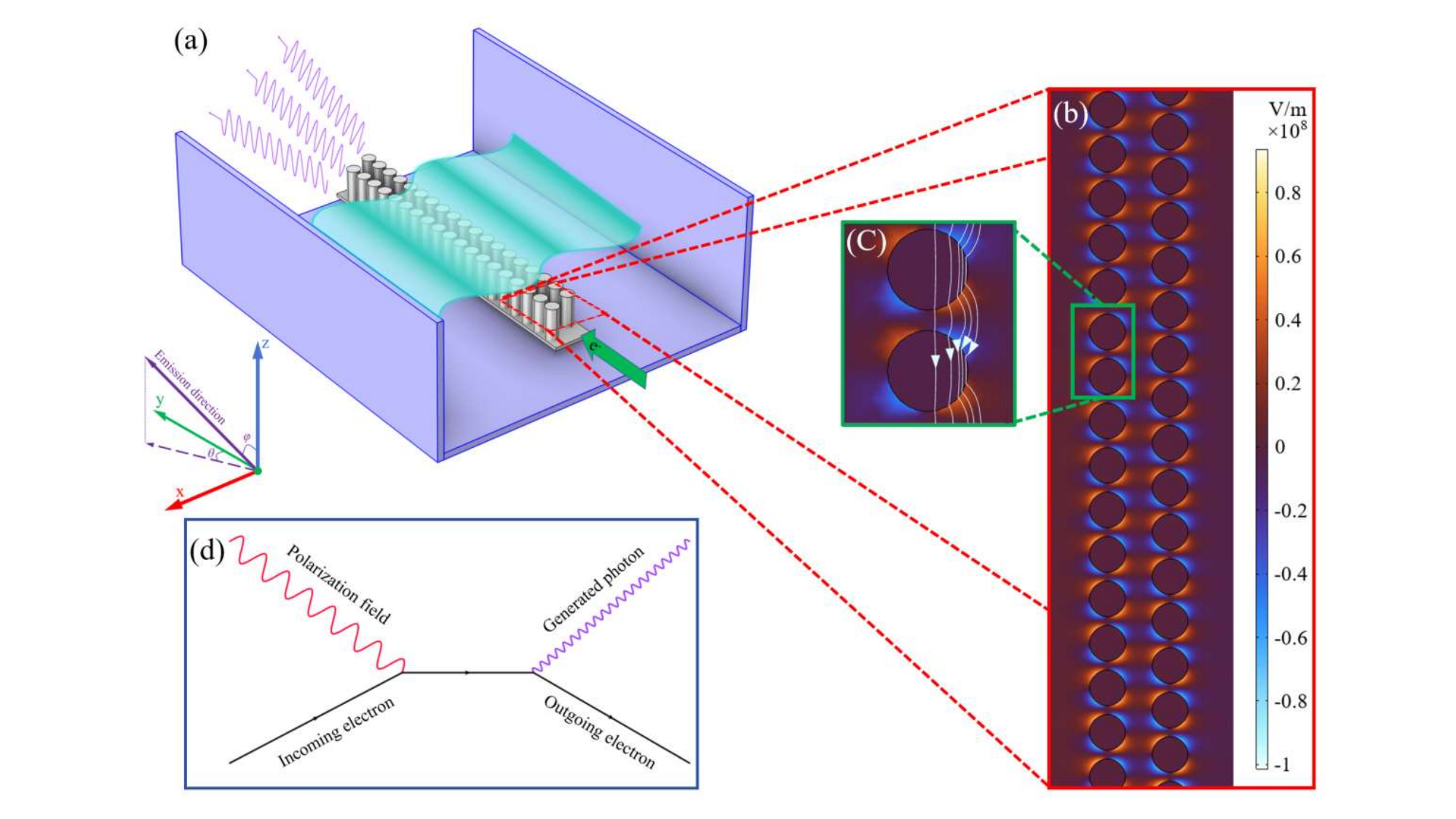}
	\caption{High energy photon radiation source based on microwave dielectric undulator. (a) The interaction between a dielectric nanopillar array and a resonant cavity (purple square box) with a microwave standing wave (green wavefront) polarized along the y-axis with an amplitude of 80 MV/m is depicted in (b), which produces a periodic transverse near field as shown in (b). (c) is the detail view of (b), which shows that the nanopillar array generates a polarization field ,with the effect of microwaves, and the white arrows show the electric field lines around the periodic near-field channel. (d) The Feynman diagram of electrons oscillating and generating radiation in this structure.}
	\label{Fig1}
\end{figure}

One of the best ways to induce small periodic oscillations of free electrons under the influence of an external field is to use electromagnetic near-field modulation. For a dielectric medium in an electric field with no surface charge, if the medium is in a vacuum environment, it is easy to obtain $(\varepsilon _1\cdot \vec{E}_1-\varepsilon _0\cdot \vec{E}_2) \cdot \vec{e}_n=0$ and $( \vec{E}_1-\vec{E}_2) \times \vec{e}_n=0$ from Gaussian theorem, where $\varepsilon_0$ and $\varepsilon_1$ are the vacuum’s and the dielectric’s dielectric constant, $\vec{E}_1$ and $\vec{E}_2$ are the electric fields inside and outside the dielectric, and $\vec{e}_n$ is the normal vector of the interface. It can be seen that if a negative y-axis electric field is applied to the dielectric cylindrical array in vacuum [see FIG. \ref{Fig1} (a)], electric field distortion will occur near the cylinder [see FIG. \ref{Fig1} (b)]. This is because the dielectric will generate displacement polarization under the effect of an external electric field, and dipole enhancement will also occur between array elements. The most noteworthy aspect is that the induced electric field of the dielectric column contains an $x$-component electric field which is not included in the external electric field [red and blue blocks in FIGs. \ref{Fig1} (b) and (c)].

By using micro/nano-processing technology, dielectric nanorods with a diameter of 40 nm are arranged in two staggered columns along the $y$-axis with a center spacing of 50 nm, resulting in a $x$-component periodic staggered near-field channel along the $y$-axis [see FIGs. \ref{Fig1} (a) and (b)]. This near-field channel will impose a periodic transverse oscillation electric field force on the injected electrons, which can make them oscillate with extremely short periods equal to 50 nm, then radiating short-wavelength photons.

In order to apply an electric field in the $y$-axis on a dielectric nanopillar array, we cannot simply use a static electric field, because a high-voltage static electric field can easily breakdown the medium. So we use the radio frequency (RF) microwave standing wave to induce near field. By placing the dielectric nanopillars at specific positions in the RF cavity, the polarization of the microwave is parallel to the $y$-axis, the channel is at the antinode of the electric wave and the nodal position of the magnetic wave, in order to generate near field which will drive electrons lateral oscillate and minimize the influence of the background magnetic field as well [see FIG. \ref{Fig2} (b)]. When electromagnetic wave passes through an array of nanopillars which orthogonal to its polarization direction, if the nanopillars’ diameter and spacing period are far less than the wavelength, the electromagnetic wave will bypass the arry due to diffraction effects and do not hinder its propagation. Therefore, it can form a stable standing wave field in this structure. In addition, the nano dielectric materials’ low absorption rate of RF photons makes them less susceptible to be damaged under high-energy microwave heating.

The displacement polarization of bound electrons in a dielectric requires a certain response time, which is generally in the range of femtoseconds to sub-nanoseconds, much shorter than the oscillation period of RF microwave. Therefore, this near field is relatively stable. The RF cavity of an electron induction accelerator can not sustain an electromagnetic field which strength is higher than the GV/m level \cite{17,18}. We hope that electrons will experience a near-field electric field strength of 10 MV/m on the $x$-axis in the oscillation channel. If the nanopillar array is made of silicon, the standing wave amplitude should be about 80 MV/m. This is feasible in engineering and there is no risk of thermal damage in the short term. If dielectric ceramic materials with higher dielectric coefficients and higher thermal damage threshold are used, the strength of the standing wave can be further reduced while the near-field electric field strength remains unchanged, or a stronger near field can be generated by using a stronger driving field to enhance the power of the light source.

It is worth noting that when the direction of the microwave electric field is reversed, the direction of the induced near field will also be reversed, which will cause the oscillation of electrons a sudden change with the $\pi$ phase. This is quite unfavorable for the monochromaticity of electron radiation. Therefore, considering that the microwave resonant cavity and microwave source of RF accelerators widely use S-band microwave (with a wavelength of about 10 cm), the length of the near-field channel of the device should be less than 5 cm, so that the direction of the electric field remains unchanged when electrons travel through the channel. In fact, in order to ensure that the driving near field intensity is sufficiently high, the near-field channel’s length is preferably less than 1 cm, allowing electrons to be injected into the device with the highest microwave electric field.

Electrons are subjected to both transverse near-field oscillations and longitudinal acceleration driven by the background microwave electric field. But, for relativistic electrons, the total increment of their longitudinal velocity when passing through the oscillation channel does not exceed the order of 10$^5$ m/s, which can be ignored compared to the longitudinal velocity close to the speed of light and the corresponding energy gain is not higher than the energy dissipation of the high-quality electron beam \cite{19}. Moreover, there is no necessary precise phase-locked condition between electrons and electric field in this model, such small changes in longitudinal velocity will not have a significant impact on the radiation output.

In addition, due to the long RF wavelength, comparing with the length of the near-field channel, the near-field can be regarded as an approximate electrostatic field when the relativistic electrons undergo several oscillation cycles, which is used in the mathematical derivation in the following text. 

\section{ELECTRONIC OSCILLATION RADIATION MODEL}
\begin{figure}[h]
	\setlength{\abovecaptionskip}{0.3cm}
	\centering\includegraphics[width=0.9\linewidth]{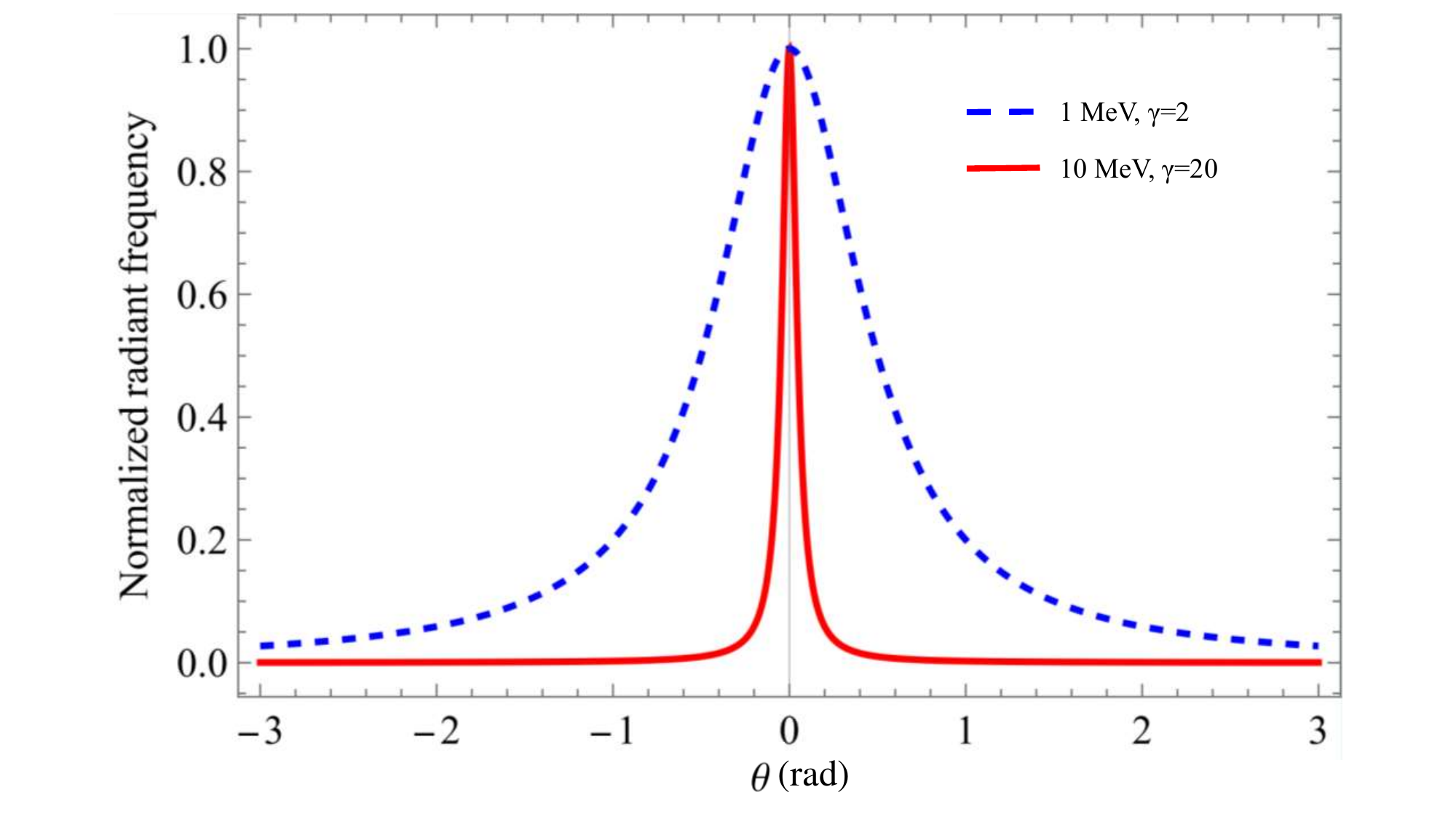}
	\caption{Dependence of electron radiation on azimuthal angle $\theta$ for electron energies of 1 MeV and 10 MeV.}
	\label{Fig2}
\end{figure}

According to the Lienart-Wiechart potential, the instantaneous electric field strength of a moving electron at the observation point with distance $R$ can be expressed as (Gaussian unit) \cite{20},
\begin{equation}
	\vec{E}\left( \vec{x},t \right) =e\left[ \frac{\vec{n}-\vec{\beta}}{k^3\gamma ^2R^2} \right] _{\text{ret}}+\frac{e}{c}\left[ \frac{\vec{n}\times \left( \vec{n}-\vec{\beta} \right) \times \dot{\vec{\beta}}}{k^3R} \right] _{\text{ret}}\label{Eq1}.
\end{equation}
Where $e$ is elementary charge, $\vec{n}$ is the unit vector from the electron to the observation point, $\vec{\beta}$ is the normalized velocity of the electron, $k=1-\vec{n}\cdot \vec{\beta}\left( t' \right) =\text{d}t/\text{d}t'$ is the wave vector, $\gamma =\sqrt{1-\beta ^2}$ is the Lorentz factor, $[]_\textbf{ret}$ is the retarded bracket, and $c$ is the light speed in the vacuum. Only the second item can radiate to a distance. It can be seen that when electrons are subjected to external forces, they will emit electromagnetic waves transmitting perpendicular to the direction of the force. 

When a relativistic electron traveling in a straight channel line under an extremely short period of transverse reciprocating electric field, it can be seen from Eq. (\ref{Eq1}) that this kind of electric field, with extremely short period and not very high intensity, will cause the electron to apply a periodic oscillating electric field (i.e. radiate electromagnetic waves) to the forward observation point, and the electron's trajectory will hardly produce significant deflection. 

In this case, $\vec{\beta}$ approaches 1 and has a direction similar to $\vec{n}$, i.e $\vec{\beta}\sim \vec{n}$, and $k=1-\vec{n}\cdot \vec{\beta}(t') \approx 1/( 2\gamma ^2 ) $, $\vec n - \vec \beta  \approx \vec n(1 - 1 + {1 \over {2{\gamma ^2}}}) = {1 \over {2{\gamma ^2}}}\vec n$. By substituting the second term of Eq. (\ref{Eq1}), the electric field of the radiation field can be obtained,
\begin{equation}
	\vec E(\vec x,t) = {e \over c}{\left[ {{{\vec n \times ({1 \over {2{\gamma ^2}}}\vec n \times \dot \vec \beta )} \over {{{({1 \over {2{\gamma ^2}}})}^3}R}}} \right]_{{\rm{ret}}}} = {e \over c}{\left[ {{{\vec n \times (\vec n \times \dot \vec \beta )} \over R}4{\gamma ^4}} \right]_{{\rm{ret}}}} \label{Eq2},
\end{equation}
the corresponding magnetic field as,
\begin{equation}
	\vec B(\vec x,t) = {\left[ {\vec n \times \vec E(\vec x,t)} \right]_{{\rm{ret}}}} \label{Eq3},
\end{equation}
and the Poynting vector as,
\begin{equation}
	\vec S(\vec x,t) = {c \over {4\pi }}{\left[ {\vec E(\vec x,t) \times \vec B(\vec x,t)} \right]_{{\rm{ret}}}} = {c \over {4\pi }}{\left[ {{{\left| {\vec E} \right|}^2}\vec n} \right]_{{\rm{ret}}}} \label{Eq4}.
\end{equation}
Observing at a small angle near the axis of electronic motion, the radiation power per unit solid angle is: 
\begin{equation}
	\begin{split}
	{{dP\left( t \right)} \over {d\Omega }} &= {c \over {4\pi }}{{\rm{[}}{R^2}{\left| {\vec E} \right|^2}{\rm{]}}_{{\rm{ret}}}} = {{{e^2}} \over {4\pi c}} \cdot 4{\gamma ^4} \cdot {\left| {\vec n \times \left( {\vec n \times \dot \vec \beta \left( t \right)} \right)} \right|^2} \\
	&= {{{e^2}{\gamma ^4}} \over {\pi c}}{\left| {\vec n \times \left( {\vec n \times \dot \vec \beta \left( t \right)} \right)} \right|^2} \label{Eq5}.
	\end{split}
\end{equation}
For relativistic electrons, there is a conversion relationship between the electromagnetic wave frequency $\nu \left( \theta  \right)$ radiated by them and the electron oscillation frequency ${\nu _0}$ on their oscillation plane ($x$-$y$ plane) \cite{21}: $\nu \left( \theta  \right) \approx 2{\gamma ^2}{\nu _0}/(1 + {\gamma ^2}{\theta ^2})$. That is, the relativistic frequency up-conversion relationship [see FIG. \ref{Fig2}]. Among them, $\theta$ is the angle between the radiation direction and the direction of electron motion, which can be approximated as the angle between the radiation direction and the y-axis in this model [see FIG. \ref{Fig1} (a)]. For relativistic electrons with higher energy, their radiation frequency has a weak dependence on the radiation direction and the angle between the $z$-axis, so can be approximated as independent.

If electric polarization near field and photon number are used to describe the power, according to the up-conversion relationship, Eq. (\ref{Eq5}) will be rewritten as:
\begin{equation}
	\begin{split}
		{{dN\left( t \right)} \over {d\Omega }} &= {{{e^2}{\gamma ^4}} \over {\pi {c^2}}}{\left| {\vec n \times \left( {\vec n \times {{{{\vec E}_P}} \over {\gamma {m_e}}}} \right)} \right|^2}{1 \over {2{\gamma ^2}h{\nu _0}}} \\
		&= {{{e^2}} \over {2\pi {c^2}h{\nu _0}}}{\left| {\vec n \times \left( {\vec n \times {{{{\vec E}_P}} \over {{m_e}}}} \right)} \right|^2} \label{Eq6}.
	\end{split}
\end{equation}
Where ${\vec E_P}$ is the polarization field, and $m_e$ is the electron mass. It can be seen that for relativistic electrons with $\gamma \gg 1$, the number of photons radiated per unit time and unit solid angle is independent of the electron energy [FIG. \ref{Fig1} (d) is a schematic diagram of the radiation generated by this model]. 

\section{RADIATION SPECTRA OF ELECTRONS WITH DIFFERENT ENERGIES IN THE UNDULATOR}
In order to generate electromagnetic waves with shorter wavelengths, the energy of electrons should be high enough to generate strong relativistic frequency up-conversion effects. However, the emissivity of high-energy electron beams is high, making it difficult to pass through channels with a width of only about 30 nm without colliding with nanorods, which will cause ionization. Therefore, we will mainly discuss low-energy relativistic electrons with energies of 1$\sim$6 MeV and high-energy electrons with energies of 10$\sim$20 MeV. The lower energy electrons can be supplied from compact accelerators, RF electron guns, even on-chip electron sources, which can be easily obtained in laboratory. 

\begin{figure}[h]
	\setlength{\abovecaptionskip}{0.3cm}
	\centering\includegraphics[width=1\linewidth]{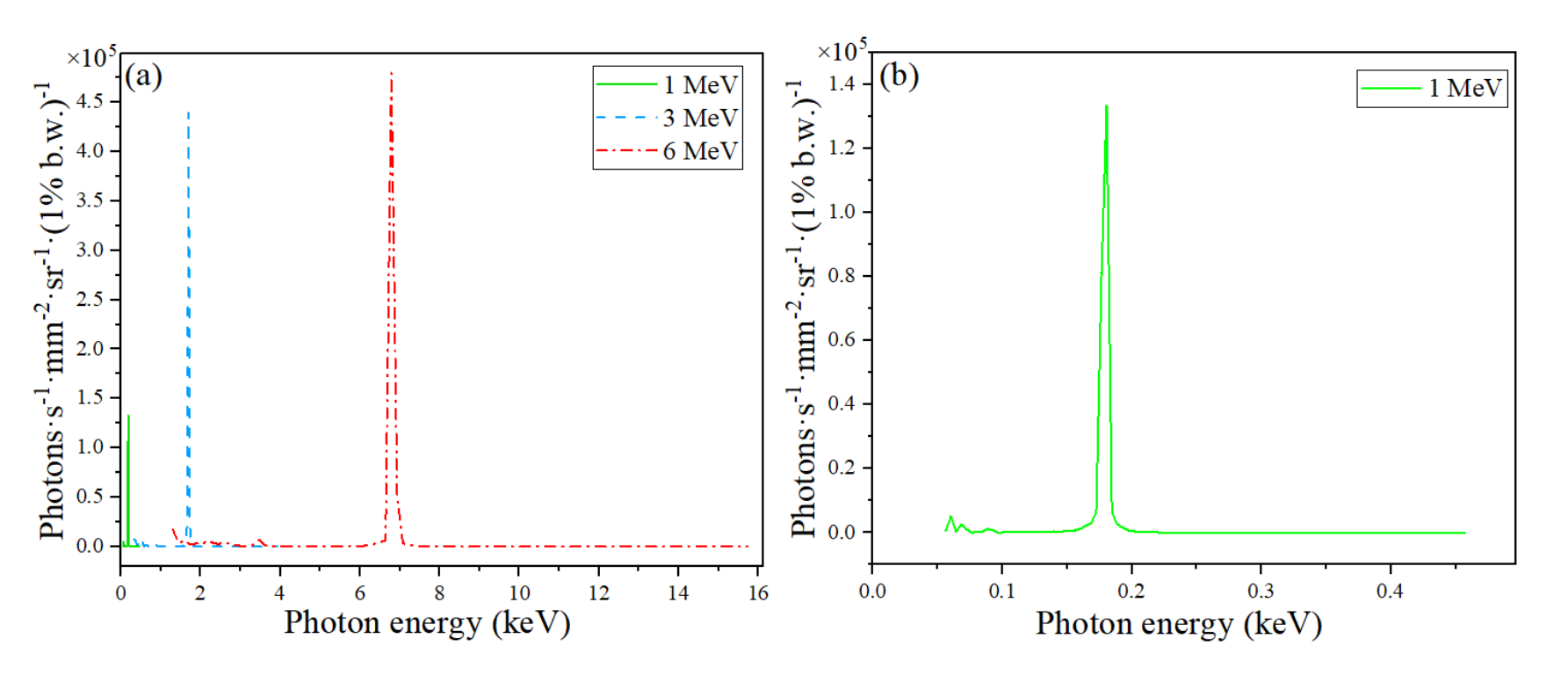}
	\caption{(a) The radiation spectra of single 1 MeV, 3 MeV, and 6 MeV electron in the device. (b) The amplification of the 1 MeV electron radiation spectrum in (a).}
	\label{Fig3}
\end{figure}

Due to the fact that the photon momentum of low-energy electron's radiation is much lower than the electron's own momentum, radiation damping can be disregarded. Use the particle tracking module of the finite element analysis (FEA) program to simulate the motion of electrons and inputs their acceleration into the Lienart-Wiechart potential calculation to obtain the electron radiation intensity. In this simulation, the device is composed of silicon columns with a diameter of 40 nm, and generates a near field of $10^8$ V/m under the drive of a 10 cm wavelength RF microwave. In common experimental environments, electron beams have a certain degree of energy dissipation, usually 1\% or even lower, which can cause spectral broadening. However, for this model, the oscillation period of relativistic electrons is fixed, and their radiation photon frequency $ \propto {\gamma ^2}$, while the energy of relativistic electrons $ \propto \gamma $. Therefore, the spectral broadening effect caused by electron dispersion should be less than 0.01\% and can be ignored.

The photon energy of lower energy relativistic electron radiation is in the range of 0.1$\sim$10 keV [see FIG. \ref{Fig3}]. The photon energy of 1 MeV electron radiation is about 0.2 keV, belonging to the extreme ultraviolet region [see FIG. \ref{Fig3} (b)], while the photons of 3 MeV and 6 MeV electron radiation are in the soft X-ray and X-ray region. From FIG. \ref{Fig3} (a), it can be observed that the number of photons radiated by 1 MeV electron per unit of time per unit of solid angle is much lower than that of 3, 6 MeV electrons, which deviates from the results of Eq. (\ref{Eq6}) to a certain extent. This is because the approximation mentioned earlier was made when the Lorentz factor $\gamma \gg 1$, which is not accurate for lower energy electrons.

In order to obtain radiation with shorter wavelengths, we can use higher energy electron beams, but correspondingly, higher energy electron beams have higher electron emissivity and it is more difficult to compress the beam diameters. Due to the fact that nanorods can extend along the $z$-axis, elliptical electron beams compressed along the $x$ direction can be used, but this process is also very difficult for extreme relativistic electrons. In order to reduce the possibility of ionization caused by electron beams colliding with nanorods, the current intensity must also significantly decrease, which means that the number of photons generated per unit time will also decrease. Therefore, an electron beam of 10$\sim$ 20 MeV is an appropriate choice for generating hard X-band.

\begin{figure}[h]
	\setlength{\abovecaptionskip}{0.3cm}
	\centering\includegraphics[width=1\linewidth]{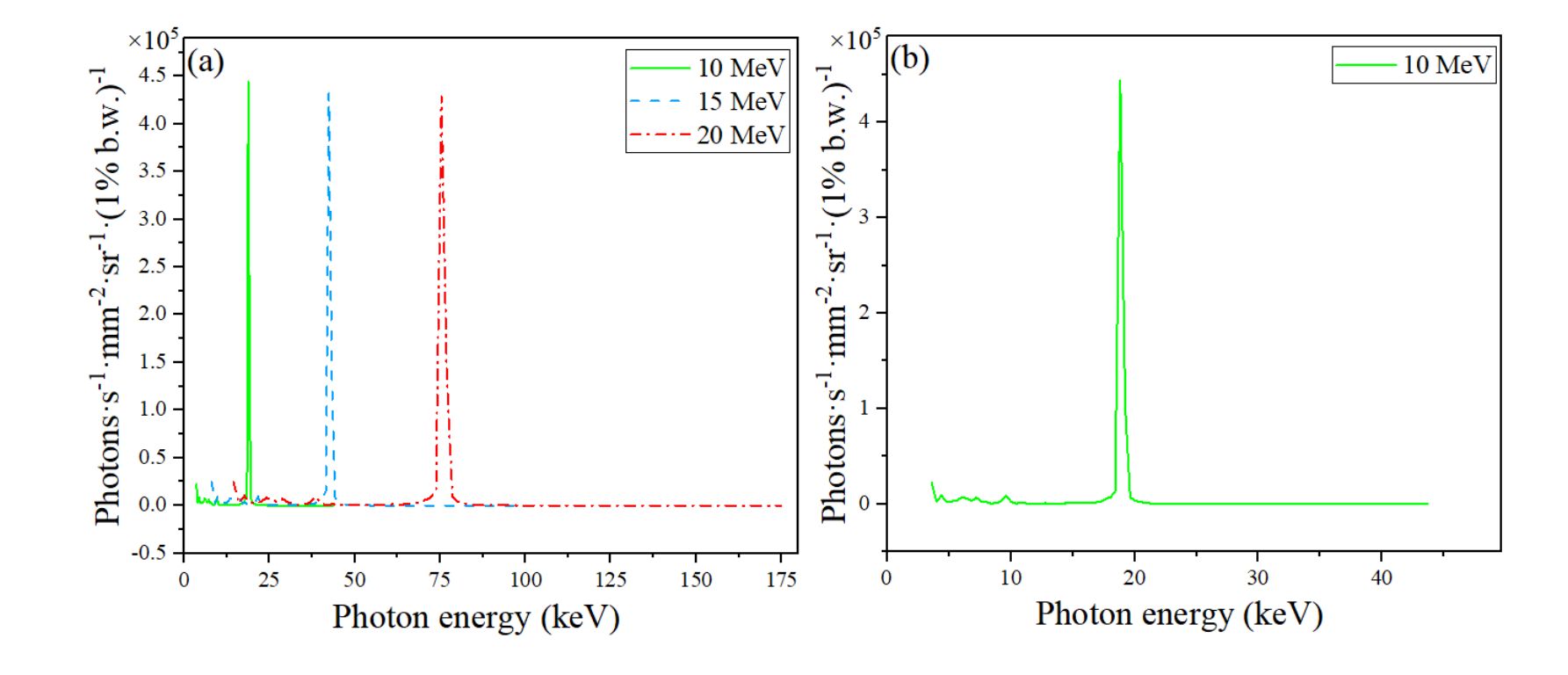}
	\caption{(a) The radiation spectra of single 10 MeV, 15 MeV, and 20 MeV electrons in the device. (b) The amplification of the 10 MeV electron radiation spectrum in (a).}
	\label{Fig4}
\end{figure}

FIG. \ref{Fig4} shows the single electron radiation intensity of 10, 15, 20 MeV electrons. It can be seen that for electrons exceeding 10 MeV, their radiation spectrum is in the hard X-Ray region. FIG. \ref{Fig4} (a) shows that the number of photons per unit time and per unit solid angle emitted by strongly relativistic electrons with different energies is approximately equal, consistent with that given by Eq. (\ref{Eq6}).

\begin{figure}[h]
	\setlength{\abovecaptionskip}{0.3cm}
	\centering\includegraphics[width=0.8\linewidth]{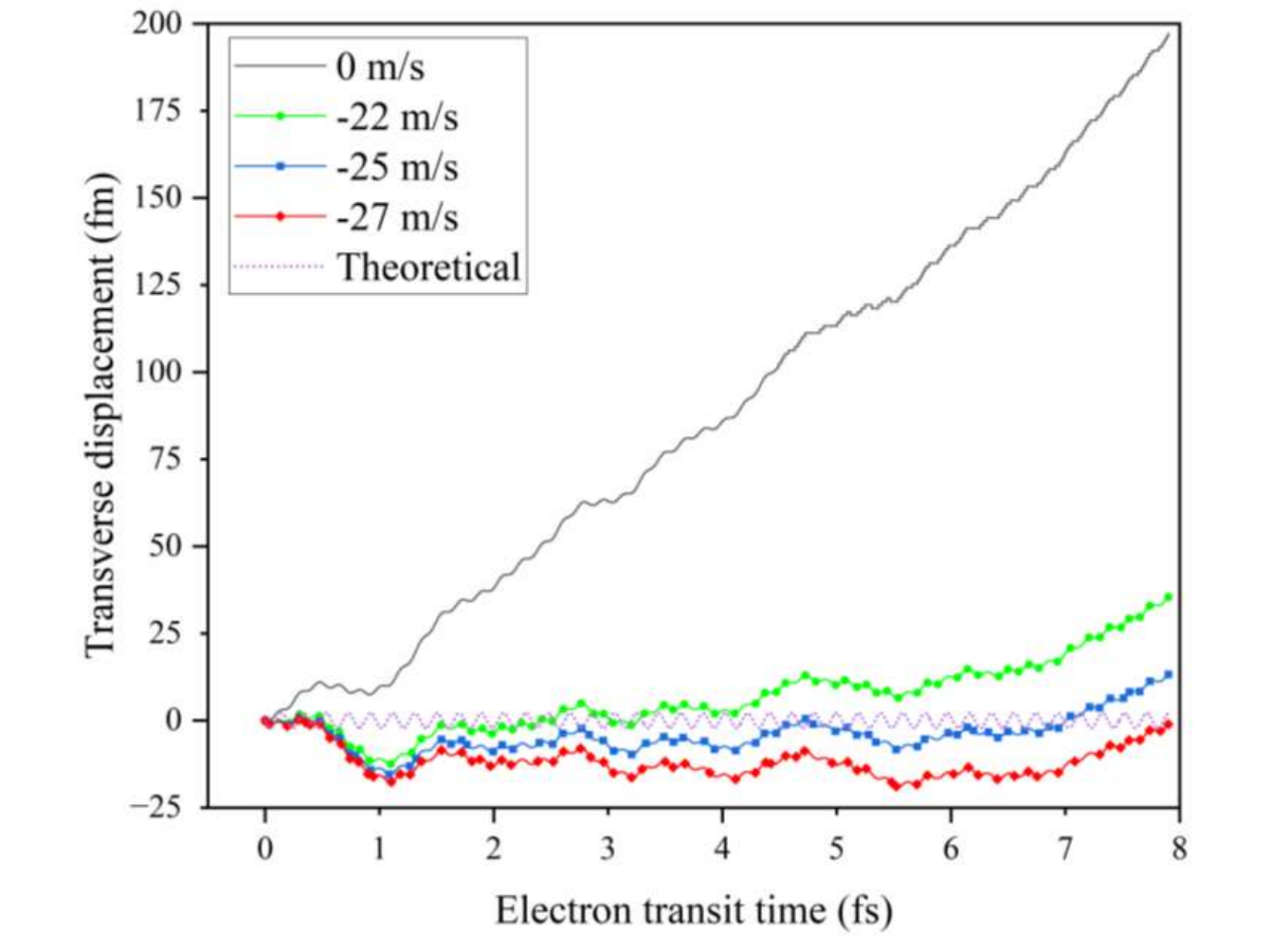}
	\caption{Transverse displacement of 1 MeV electron within 50 cycles. The figure shows the electron’s trajectory without initial lateral velocity modulation, the trajectory with initial lateral velocity modulation, and the theoretical motion trajectory. It can be seen that the lateral displacement of electrons is very small, and the initial lateral velocity modulation can greatly reduce the electron's walk off.}
	\label{Fig5}
\end{figure}

Combining FIGs. \ref{Fig3} and \ref{Fig4}, it can be seen that the electromagnetic wave spectrum radiated by a single energy electron beam in this device is a quasi-monochromatic spectrum with excellent monochromaticity. As mentioned earlier, compared with traditional FEL systems, the transverse displacement and the changes of velocity direction of electrons in this system are very small. FIG. \ref{Fig5} shows the transverse displacement of a 1 MeV electron during approximately 50 oscillation cycles, only at the femtometer level. For higher energy electrons, due to the increase in dynamic mass, their transverse displacement comes to the sub-femtometer level. Due to the relaxation process of the dipole polarization field and dipole enhancement generated by dielectric materials, the oscillation trajectory of electrons under the polarization field is not a standard sine curve [purple dashed line in FIG. \ref{Fig5}], and often moves away to one side. If an appropriate transverse velocity is given to the electrons before they are injected into the undulator (such as through a parallel charged electrode plate), the walk-off effect can be greatly reduced, further compressing the transverse motion range of the electron. 

The extremely small transverse oscillation amplitude causes the undulator factor of this device K$\sim$0 \cite{22}, resulting in a very small divergence angle of radiation, which can be approximated as $1/\gamma$. Therefore, one can say that the higher energy the electron has, the smaller the divergence angle of radiation is. The output area of the device is only on the order of $\mu$m$^2$, so the radiation brightness is quite high.

\section{CONCLUSION}\label{summary}
We propose for the first time a mechanism based on the interaction between microwave and nanostructures to generate periodic oscillations of relativistic electrons in the near field, promoting the generation of extremely short wavelength electromagnetic radiation. The melting point of dielectric materials such as monocrystalline silicon and dielectric ceramics is mostly above 1000 $^{\circ}$C and have low absorption rate of RF photons, which make this configuration is less prone to thermal damage during operation.

Because the oscillation period of electrons in this model is entirely determined by the period of nanopillar array, high brightness X-ray radiation with excellent monochromaticity and minimal divergence angle can be generated by using an RF photocathode electron gun or a desktop accelerator system. Moreover, due to the many similarities between this structure and the structure of the Laser Dielectric Acceleration (DLA) device, both using an external driving electromagnetic wave field to irradiate periodic nanostructures and generate near-field modulation of electron motion. Therefore, after adjusting the period and arrangement of the nanostructure, higher energy laser irradiation can also be used to apply for higher acceleration to electrons in this device, thereby increasing the power of electron oscillation radiation. Furthermore, this structure can be etched on the same substrate with the DLA system, allowing the acceleration, regulation, and oscillation radiation processes of the electron beam to be completed on the same chip, within a distance of a few centimeters.

If laser is used instead of microwave as the driving source in the future, the electromagnetic damage threshold of the dielectric material itself is much higher than that of the metal resonant cavity. Therefore, lasers with higher electric field intensity can be used to generate stronger near field. This means that the power of electron radiation can be increased by several orders, or the width of the oscillation channel can be greatly expanded while maintaining the force intensity of electrons on the central axis. This means that extreme relativistic electrons with higher emissivity can also be injected into it to generate $\gamma$-ray or even hard $\gamma$-ray.

\emph{Acknowledgments:} This work has been supported by National Natural Science Foundation of China (No. 12388102, 12074398, 12325409, U226720057); CAS Project for Young Scientists in Basic Research (YSBR-060) and Shanghai Pilot Program for Basic Research, Chinese Academy of Sciences, Shanghai Branch.

\end{document}